\documentclass{article}
\usepackage[a4paper,top=3cm,bottom=3cm,left=2.5cm,right=2.5cm,marginparwidth=3cm]{geometry}
\usepackage[utf8]{inputenc}
\usepackage[super,comma,numbers,sort&compress]{natbib}
\usepackage{amsmath}
\usepackage{graphicx}
\usepackage{listings}
\usepackage{comment}
\usepackage{authblk}
\usepackage{enumitem}
\usepackage{multirow}
\usepackage{url}

\title{Free Energy Landscapes, Diffusion Coefficients, and Kinetic Rates from Transition Paths}

\author[1]{Karen Palacio-Rodriguez}
\author[1]{Fabio Pietrucci\footnote{fabio.pietrucci$@$sorbonne-universite.fr}}
\affil[1]{\small{Mus\'eum National d'Histoire Naturelle, UMR CNRS 7590, Institut de Min\'eralogie, de Physique des Mat\'eriaux et de Cosmochimie, IMPMC, Sorbonne Universit\'e, F-75005 Paris, France}}

\date{}

\begin{document}

\maketitle

\begin{abstract}
We address the problem of constructing accurate mathematical models of the dynamics of complex systems projected on a collective variable. To this aim we introduce a conceptually simple yet effective algorithm for estimating the parameters of Langevin and Fokker-Planck equations from a set of short, possibly out-of-equilibrium molecular dynamics trajectories, obtained for instance from transition path sampling or as relaxation from high free-energy configurations. The approach maximizes the model likelihood based on any explicit expression of the short-time propagator, hence it can be applied to different evolution equations. We demonstrate the numerical efficiency and robustness of the algorithm on model systems, and we apply it to reconstruct the projected dynamics of pairs of C$_{60}$ and C$_{240}$ fullerene molecules in explicit water. Our methodology allows reconstructing the accurate thermodynamics and kinetics of activated processes, namely free energy landscapes, diffusion coefficients, and kinetic rates. Compared to existing enhanced sampling methods, we directly exploit short unbiased trajectories, at a competitive computational cost.
\end{abstract}

\section*{Introduction}

Atomistic computer simulations of rare events like phase transitions, chemical reactions, or biomolecular interactions and conformational changes have three paramount goals: predicting detailed mechanisms, free energy landscapes, and kinetic rates. All of these tasks are, in many cases, cumbersome and require intensive human and computer effort,  the calculation of rates being the most difficult.~\cite{Pietrucci17review,Camilloni18} 

Usually, practitioners adopt the following strategy: first, a guess of the transformation mechanism is obtained using chemical intuition or acceleration techniques including artificial biasing forces. Next, either the free energy landscape as a function of a few collective variables (CVs) is reconstructed, or the detailed mechanism is investigated with transition path sampling (TPS) techniques \cite{Dellago08,Bolhuis21}. Alternative approaches like transition interface sampling \cite{VanErp05} or forward flux sampling \cite{Allen05} seek at the same time mechanism, free energy, and kinetic rates; however, their computational cost can be prohibitive and the results could depend on the choice of the order parameter required by these techniques. 

Projecting the dynamics of the system on one (or a few) CV $q$ leads to a theoretical framework describing equilibrium thermodynamics through free energy landscapes, and kinetics (including out-of-equilibrium relaxation) through Langevin and Fokker-Planck equations \cite{Zwanzig01,Risken96}. 
Three main types of Langevin equations that depend on the nature of the physical system, and on the observational time resolution $\tau$, appear in the literature.
In the non-Markovian case, a generalized Langevin equation is customarily employed,
including a deterministic force (the gradient of the free energy $-dF/dq$) and  time-correlated friction and noise, connected by the fluctuation-dissipation theorem. \cite{Zwanzig01,Luczka05} 
Such time correlation, or memory effects, can be significant for the short-time dynamics of small solutes immersed in a liquid bath \cite{Grote80,Lee15,Daldrop18}.

However, in many applications in physics, chemistry or biology an observational time resolution $\tau$ coarser than the memory time scale is pertinent, so that a Markovian Langevin equation accurately describes the projected dynamics. Depending on the intensity of the friction, or, in other words, on the extent of its characteristic time compared to $\tau$, which depends on the physical process and on the choice of CVs, a second-order underdamped Langevin equation featuring inertia (and reminiscent of Hamilton's equations) or a first-order overdamped equation is appropriate \cite{VanKampen92,Coffey12,Lu14}. 

A number of algorithms aim at parametrizing Langevin models starting from  trajectories of many-particle systems \cite{Straub87,Tuckerman93,Timmer00,Gradivsek00,Chorin02,Hummer03,Hummer05,Best06,Lange06,Horenko07,Darve09,Micheletti08,Berezhkovskii11,Zhang11,Crommelin11,Crommelin12,Lee15,Schaudinnus15,schaudinnus16,Lesnicki16,Meloni16,Daldrop18,Biswas18,Freitas19,Nuske19,Baldovin20,Lickert20,Wang20,Ayaz21,Vroylandt22}.
More in detail, non-Markovian friction (the so-called memory kernel) can be reconstructed in different ways, for instance based on a set of time-correlation functions and Volterra integral equations (see, e.g., refs.~\citenum{Tuckerman93,Lee15,Daldrop18}) or via likelihood maximization~\cite{Vroylandt22}. In the case of Markovian (overdamped) Langevin equations, the latter approach~\cite{Hummer05,Micheletti08,Zhang11}, together with direct estimation of Kramers-Moyal coefficients~\cite{Zhang11,Meloni16} have been employed.

Often, only friction is addressed, assuming that the free-energy profile is known in advance, or that it can be directly estimated from the histogram of the CV based on long equilibrium trajectories reversibly sampling all relevant states (see, e.g., refs.~\citenum{Lange06,Micheletti08,Daldrop18}). Clearly, this severely limits the scope of the techniques, given that ergodic molecular dynamics (MD) trajectories can be affordably generated only in the case of low kinetic barriers (a few $k_BT$).

In the case of rare events, some approaches are able to exploit enhanced sampling simulations, such as umbrella sampling, to alleviate the time scale problem \cite{Hummer03,Zhang11,Biswas18}, even though the cost of such simulations to estimate free-energy and diffusion profiles remains considerable in the presence of high barriers. Moreover, such acceleration techniques usually rely on an initial guess of CVs to be biased: a bad choice can lead to poor convergence and to explore unfavorable transition mechanisms compared to brute force MD, while a change of CVs requires in general generation of new biased trajectories, multiplying the costs. At variance with previous attempts, in this work we address this issue by basing our Langevin optimization approach on unbiased TPS-like~\cite{Bolhuis02} trajectories, spontaneously relaxing from barrier tops, and faithfully reproducing brute force transitions thanks to the reliability of TPS techniques. In our case, the choice of the CV is performed {\sl a posteriori}, and multiple CVs can be tested at negligible computational cost since no new MD trajectories are required.

Whether the projected dynamics of a given system can be accurately modeled by a specific flavor of Langevin equation depends upon two main observational choices: the CV $q$ and the time resolution $\tau$ of the trajectory. In particular, it is expected that memory effects decrease when CVs become similar to the committor function.~\cite{Hummer03,Lu14,Tiwary15pnas}. For a fixed CV, too small $\tau$ can introduce non-Markovian effects.
Albeit acknowledged in the literature, these principles have seldom been applied to critically assess how accurately a given Langevin model reproduces the original dynamics for different CVs and different time resolutions~\cite{Micheletti08,Meloni16,Lickert20}. In many cases a single system is studied, and Langevin predictions at fixed time resolution are compared with independent estimates of free energies and rates.
We also remark on a further important point: $\tau$ cannot be freely increased to enhance Markovian behavior, since too large values rarefy the sampled trajectory to the point of lacking information on barrier regions.

As a result of all the previous concerns and difficulties, to date CV-based Langevin equations have been used sporadically and most often heuristically, and are not systematically applied to access the thermodynamics and kinetics of a wide range of complex physicochemical processes. It is our aim to improve the state of the art in this respect, with particular care to the assessment of time resolution effects on the accuracy of the reduced-dimensionality model, as detailed below.

In this work we present a conceptually simple and computationally efficient method to construct data-driven Langevin models of rare events in complex systems. 
Since equilibrium MD data are not necessary for training, the method can be applied irrespective of the height of the barriers separating metastable states.
We also provide an explicit procedure to assess the optimal resolution $\tau$ leading to accurate Langevin models, and to predict their reliability. The work is organized as follows: $(i)$ we describe the algorithm used to optimize the models and the computational details of the MD simulations, $(ii)$ we test the algorithm on a double-well potential with known free energy, diffusion profiles and kinetic rates, $(iii)$ we apply the algorithm on the interaction of C$_{60}$ and C$_{240}$ fullerene dimers in explicit water, a proxy model for protein-protein interactions, allowing also to evaluate the effect of the barrier height. Finally, we discuss the scope of the method, its limitations and future application perspectives.

\section*{Theoretical Methods}

\subsection*{A. Optimization of Langevin Models (Parameter Estimation)}

Langevin equations can be obtained from Hamilton's equations of motion by projecting the high-dimensional deterministic dynamics on a subset of the phase-space variables \cite{Zwanzig01,Vroylandt22}.
In the limit of strong friction, velocity fluctuations away from the equilibrium distribution are damped quickly (compared to the time resolution $\tau$), yielding the widely-employed overdamped Langevin equation:

\begin{equation}\label{eq:OLEposdep}
\dot q = -\beta D(q) \frac{\partial F(q)}{\partial q} +\frac{\partial D(q)}{\partial q} +\sqrt{2 D(q)}\,\eta(t)
\end{equation}

where $F(q)=-k_BT\log\rho_\mathrm{eq}(q)$ is the free-energy landscape, $D(q)$ is the position-dependent diffusion coefficient, and $\eta(t)$ is a Gaussian white noise. We remark that a realistic description of the dynamics generally requires a nonconstant $D(q)$ \cite{Hummer05}, and that the use of the overdamped equation can be formally justified (for instance, it provides exact mean first passage times (MFPT)) when $q$ is the optimal CV, that is, any monotonic one-to-one function of the committor \cite{Lu14}. 

The corresponding Fokker-Planck equation is the Smoluchowski equation \cite{Carmeli83}, describing the evolution of the probability density $\rho(q,t)$ from an initial distribution $\rho(q,0)$ (typically $=\delta(q-q_0)$) under given boundary conditions:

\begin{equation}\label{eq:Smoluchowski}
\begin{split}
\frac{\partial}{\partial t}\rho &=\frac{\partial}{\partial q}\left( 
\beta D\frac{\partial F}{\partial q}\rho -\frac{\partial D}{\partial q}\rho \right)
+\frac{\partial^2}{\partial q^2}\left(D\rho\right) \\
&=\frac{\partial}{\partial q}\left[ D e^{-\beta F} 
\frac{\partial}{\partial q} \left( e^{\beta F} \rho \right) \right]
\end{split}
\end{equation}

To construct an optimal model of the $q$-projected dynamics, we adopt an efficient algorithm that maximizes the likelihood $\mathcal{L}(\mathbf{\theta})$ to observe the trajectory $\{q_i\equiv q(t_i)\}_{i=1,...,N}$, $t_{i+1}-t_i=\tau$ given the Langevin model eq \ref{eq:OLEposdep}. The parameters $\mathbf{\theta}$ encode the shape of $F(q)$ and $D(q)$. To obtain an explicit form for the likelihood, we consider the formal solution of the Fokker-Planck equation
$\partial_t\rho=L\rho$ (eq \ref{eq:Smoluchowski}), calling $L$ the Fokker-Planck operator (the adjoint of the generator of the stochastic process) \cite{Risken96,Leimkuhler16}:

\begin{equation}\label{eq:formal_solution}
\begin{split}
p(q',t+\tau | q,t) & = e^{L\tau} \delta(q'-q) \\
&= \left[ 1 + L\tau + \frac{1}{2}L^2\tau^2 + ... \right] \delta(q'-q)
\end{split}
\end{equation}

For small $\tau$, the short-time transition probability density $p$, also called propagator, is an ingredient of path-integral formulations of long-time transition probabilities, and, neglecting terms of the order of $\tau^2$, can be explicitly obtained (using Fourier transforms \cite{Risken96,Zhang11}) in Gaussian form:

\begin{equation}\label{eq:prop}
p(q',t+\tau | q,t) \approx \frac{1}{\sqrt{2\pi \mu}}
e^{-(q'-q-\phi)^2/2\mu}
\end{equation}

\begin{equation}\label{eq:param}
\phi = a\tau = (-\beta D F' + D')\tau
\ \ ,\ \mu = 2D\tau
\end{equation}

where the prime indicates $d/dq$, and $a$ is the drift term in eq \ref{eq:OLEposdep} (note that this propagator is exact for any $\tau>0$ if $F'$ and $D$ are constant). A better approximation, still retaining the Gaussian form eq \ref{eq:prop}, can be obtained by keeping terms up to order $\tau^2$ in eq \ref{eq:formal_solution}, applying for instance Drozdov's approach based on the cumulant-generating function: \cite{Drozdov97}

\begin{equation}\label{eq:prop2}
\phi = a\tau +\frac{1}{2}(a a' + D a'')\tau^2
\ \ ,\ \mu = 2D\tau +
(aD' + 2a'D + DD'') \tau^2
\end{equation}

Armed with this explicit expression for the short-time propagator, it is possible to write the likelihood of the observed trajectory $\{q_i\}_{i=1,...,N}$:

\begin{equation}\label{eq:logL}
-\log \mathcal{L}(\theta)=
\sum_{i=1}^{N-1} \frac{1}{2}\log[2\pi \mu_i(\tau)] 
+\sum_{i=1}^{N-1}\frac{[q_{i+1}-q_i-\phi_i(\tau)]^2}{2\mu_i(\tau)} 
\end{equation}

where $\phi_i(\tau)\equiv\phi(q_i,\tau)$, $\mu_i(\tau)\equiv\mu(q_i,\tau)$.
The optimization of the Langevin model is achieved by minimizing $-\log \mathcal{L}(\theta)$ as a function of the parameters, directly yielding the free energy and diffusion profiles. Equation \ref{eq:logL} is minimized using a simple iterative stochastic algorithm: starting from an initial guess for the free energy and diffusion profiles (represented on a discrete grid of 1000 points), there are two types of trial moves (with probabilities $20\%$ and $80\%$, respectively): (i) adding to the profiles a Gaussian at a random position, with a random width between $(q_{max}-q_{min})/20$ and $(q_{max}-q_{min})$ and height between $-k_BT/2$ and $k_BT/2$ (with $20\%$ probability to modify either profile, and $80\%$ to modify both, at each optimization step), or (ii) scaling the profiles by a Gaussian-distributed random factor with mean 1 and standard deviation 0.3.
The moves are accepted or rejected on the basis of a Metropolis criterion on $(-\log \mathcal{L}_\mathrm{new} + \log \mathcal{L}_\mathrm{old})/\Tilde{T}$. The effective temperature $\Tilde{T}$ is automatically scaled during the iterations to keep the acceptance close to a target value of 5\%.

\begin{figure*}[ht!]
\centering
\includegraphics[width=\linewidth]{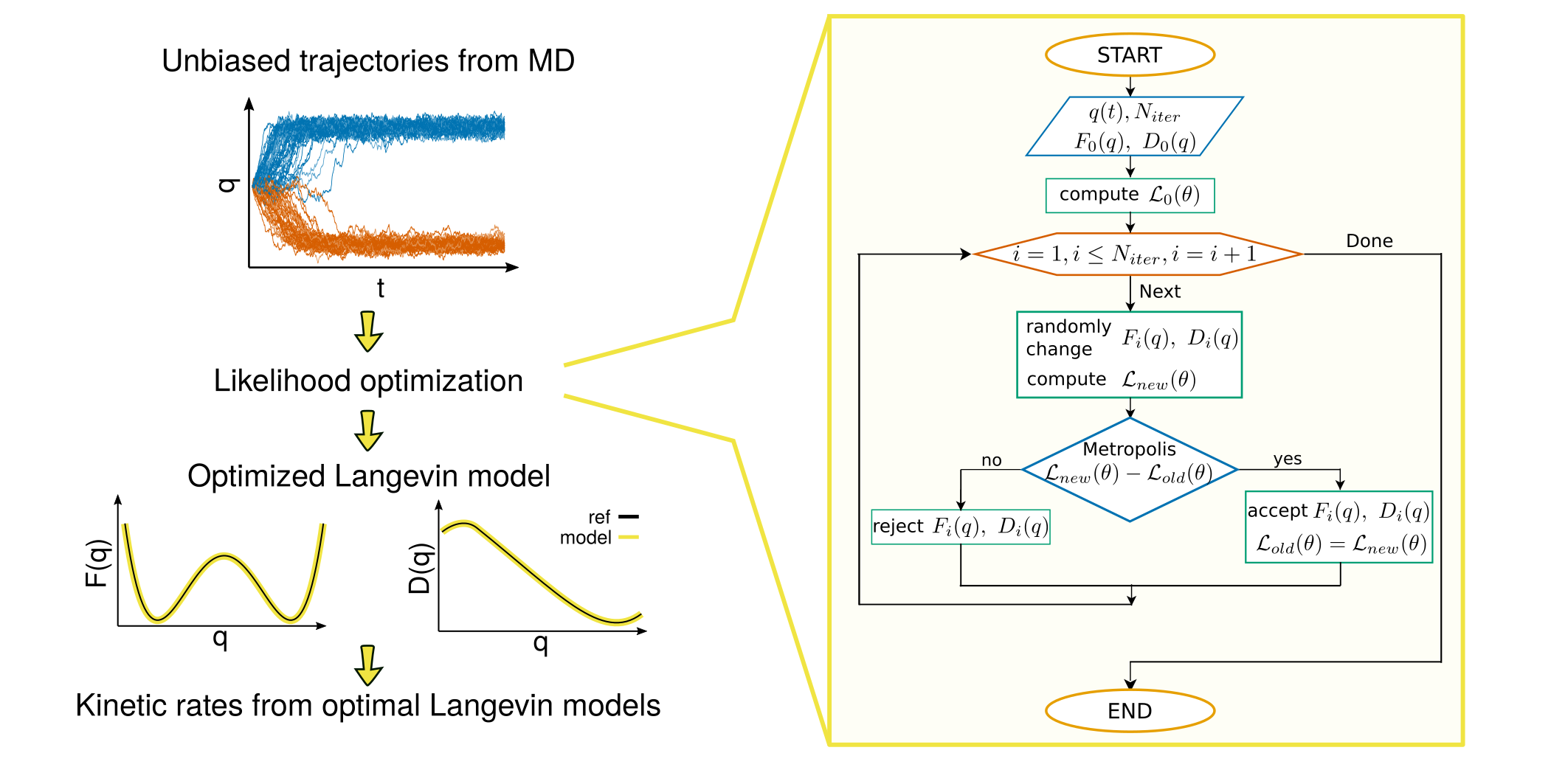}
\caption{Schematic illustration of the approach proposed to optimize Langevin models of projected dynamics.
The input data are formed by unbiased MD trajectories projected on a selected collective variable, $q(t)$. A likelihood function $\mathcal{L}(\theta)$ is maximized (i.e., $-\log \mathcal{L}(\theta)$ is minimized) by randomly varying the parameters encoding the free energy $F(q)$ and diffusion profile $D(q)$ of an overdamped Langevin equation, until optimally modeling the original MD data. At each iteration, the new parameters are accepted or rejected according to a Metropolis criterion. The optimal Langevin model corresponds to the last accepted $F(q)$ and $D(q)$ profiles, and is used for further estimation of kinetic rates (see Methods for details). 
}\label{fig:algorithm}
\end{figure*}

The algorithm  we propose for the optimization of the likelihood function 
(Figure \ref{fig:algorithm}) is straightforward:

\begin{enumerate}[itemsep=2pt,topsep=0pt,parsep=0pt,partopsep=0pt]
  \item Start with trajectories $q(t)$ from MD and with an initial guess for the free energy and diffusion profiles.
\item Compute $-\log \mathcal{L}(\theta)$ for the initial data.
\item Make a random change of the free energy and diffusion profiles and compute the new $-\log \mathcal{L}(\theta)$.
\item Use a Metropolis criterion to accept or reject the new parameters.
\item Repeat $N_\mathrm{iter}$ times the procedure.
\end{enumerate}

Note that alternatives to stochastic Monte Carlo optimization could be used; however, we appreciate the simplicity and generality of this approach, that does not require estimation of the derivatives $\partial \mathcal{L}/\partial\theta$. In all cases, we set $N_\mathrm{iter}=10^6$.

The initial guess of the parameters can facilitate the optimization of the likelihood function: without requiring any additional data besides the projected MD trajectories, an approximate value of $D$ in a free-energy minimum can be estimated from the variance of $q$ divided by its autocorrelation time~\cite{Hummer05}. As sketched in Figure \ref{fig:algorithm} the input trajectories $q(t)$ typically contain a final part fluctuating in the bottom of a well, that can be employed in the latter formula. The profile $D(q)$ is simply set equal to the constant estimated value. As for the initial $F(q)$, we found that the specific form is largely irrelevant. For the sake of simplicity we start from a flat profile $F(q)=0$.

The code and input files are freely available on GitHub (\url{https://github.com/physix-repo/optLE}).

\subsection*{B. Integration of the Langevin Equation of Motion}

To generate input data for the benchmark systems, as well as to test the quality of the approximate propagator used in the likelihood function (see previous section), we numerically integrate Langevin trajectories using the Milstein algorithm, an order 1 strong Taylor scheme superior to Euler-Maruyama in the case of multiplicative noise (i.e., position-dependent diffusion): \cite{Kloeden12}

\begin{equation}\label{eq:integrator}
\begin{split}
    \Delta q = & \left[ -\beta D(q) F'(q) +\frac{1}{2} D'(q) \right] \Delta t \\
    & +\sqrt{2 D(q) \Delta t}\, G 
    + \frac{1}{2}D'(q)\Delta t\, G^2
\end{split}
\end{equation}

where $G$ is a Gaussian-distributed random number with zero mean and unit variance.
The integration time step $\Delta t$ is chosen after testing the observed long-time histogram of $q$ against $e^{-\beta F(q)}$, where $F(q)$ is the exact landscape used in the Langevin equation.

\subsection*{C. Optimal Time Resolution of the Reduced Model}
\label{sec:diagnostics-tau}

The accuracy of the optimized Langevin model with respect to the original MD trajectory $q(t)$ crucially depends on the time resolution $\tau$ adopted,  $q(t) \equiv q(k\tau), k=1,...,N$. In fact, it is expected that only for a sufficiently large $\tau$ the overdamped equation can model accurately the MD data. To guide the choice of $\tau$ we introduce diagnostics able to predict what choices lead to accurate models.

In general, for given profiles $F(q)$ and $D(q)$, a numerical Langevin integrator can be ``inverted'' to estimate the value of an effective noise $G(t)$ corresponding to each observed displacement $\Delta q= q(t+\tau)-q(t)$ in the input MD trajectory. 
The simplest formula corresponds to the Euler-Maruyama integrator:

\begin{equation}\label{eq:effnoise-em}
  G = \frac{\Delta q + (\beta D F' - D') \tau}{\sqrt{2 D \tau}}
\end{equation}

Another option is to invert the Milstein integrator, eq \ref{eq:integrator}:

\begin{equation}\label{eq:effnoise-m}
  G = \frac{1}{D' \tau} \Bigg( -\sqrt{2D\tau} + \sqrt{(2D\tau)- (2D'\tau)\bigg( \frac{1}{2}D'\tau - \beta DF'\tau - \Delta q \bigg) }\Bigg)
\end{equation}

If the resulting effective noise would have been generated by the model, the $G(t)$ trajectory would be random, with zero mean, unit variance and no time correlation: $\langle G(t)G(t')\rangle=\delta_{t,t'}$ (the latter symbol indicating the Kronecker delta, since time is discretized). 
Therefore, to estimate the expected accuracy of the optimal Langevin model it is possible to inspect the mean, variance, and correlation time $\tau_\mathrm{noise}$ of the effective noise \cite{Micheletti08,Lickert20}. For the latter quantity, we adopt here a simple operative definition, as
the first time where the autocorrelation function $C(t)=\langle G(0)G(t)\rangle/\langle G^2\rangle$ drops below 1/100. Note that in case of oscillatory behavior of $C(t)$ it is important to inspect the latter to assess if the estimate is acceptable. We find that the expressions for the noise from Euler-Maruyama (eq \ref{eq:effnoise-em}) or Milstein (eq \ref{eq:effnoise-m}) integrators yield similar results (see Supporting Information Figure S1).

As a second metric to diagnose problems in model accuracy we generate (using the integrator eq \ref{eq:integrator}) $M=100$ Langevin trajectories of duration $\tau$ and time step $\Delta t \ll \tau$ starting from each point of the input trajectory 
$\{q_i\}_{i=1,...,N}$. The resulting final displacements $\Tilde{q}_{i+1}^k-q_i$ (with $k=1,...,M$) are shifted and scaled according to the theoretical propagator of the model, eq \ref{eq:prop}, by subtracting $\phi_i$ and dividing by $\sqrt{\mu_i}$. In this way, their distribution would be Gaussian with zero mean and unit variance if the propagator was exact, and an average likelihood can be evaluated to estimate deviations from such ideal behavior:

\begin{equation}\label{eq:Lprop}
-\overline{\log\mathcal{L}_\mathrm{prop}} = 
\frac{1}{N-1}\sum_{i=1}^{N-1} 
\frac{1}{M}\sum_{k=1}^{M} \frac{1}{2} \left[  
\log 2\pi
+\left( \frac{\Tilde{q}_{i+1}^k-q_i-\phi_i}
{\sqrt{\mu_i}} \right)^2
\right]
\end{equation}

which yields $\frac{1}{2}[\log 2\pi+1]\approx 1.419$ in the ideal case. For increasing $\tau$, the propagator becomes unreliable, reducing the likelihood and hence allowing to predict, once again, the quality of the model (see Figure \ref{fig:dw-res}, \ref{fig:ful-rates}).

\subsection*{D. Kinetic Rates}

In general cases, the kinetic rate for a transition A$\rightarrow$B can be computed  by means of the reactive flux (Bennett-Chandler) technique, that includes recrossing effects beyond transition state theory \cite{Jungblut16}
$$
k(t) = \frac{\langle \dot q(0) \, \delta[q(0)-q^*] \, h_B[q(t)]  \rangle }{ \langle h_A \rangle }
$$
where $h_A(q)$ and $h_B(q)$ are the indicator functions of metastable states $A$ and $B$.
Under the hypothesis of rare transition events, there is a plateau in the values of $k$ for a broad interval of $t$ values larger than the typical microscopic time scale (e.g., of molecular vibrations) and smaller than the MFPT. $q^*$ is any point close to the separatrix: only efficiency is affected if they differ significantly. 

The rate can be computed numerically by shooting $N_s$ MD simulations from $q^*$ with random atomic velocities drawn for the Maxwell-Boltzmann distribution: for a given $t$, $k(t)$ is the product of the sum of initial velocities for trajectories ending up in $B$, divided by the total number of shots $N_s$, and of a term containing information about the free energy barrier

\begin{equation}\label{eq:reactiveflux}
k(t) = \frac{\sum_{j=1}^{N_s} \dot q_j(0) h_B[q_j(t)]}{N_s}
\frac{e^{-\beta F(q^*)}}{\int_{\Omega_{A}} dq\,e^{-\beta F(q)} } 
\end{equation}

We remark that for rare transitions with overdamped one-dimensional dynamics it is also possible to compute the MFPT (the inverse of the rate) as an integral over the free energy and diffusion landscapes:

\begin{equation}\label{eq:mfpt-Smoluchowski}
k^{-1}(q_0) = \int_{q_0}^b \mathrm{d}y \frac{\mathrm{e}^{\beta F(y)}}{D(y)} \int_{a}^y \mathrm{d}z e^{-\beta F(z)}    
\end{equation}

where, in the typical case of the escape from a well (even though the formula is general), $q_0$ is the initial position in the well, $a$ is the position of a reflective boundary located on the opposite side as the transition barrier, and $b$ is the position of an absorbing boundary located beyond the barrier.~\cite{Jungblut16} 
In the following, we adopt as location of the absorbing boundary $b$  the first minimum in ($\frac{\mathrm{d}}{dq})k^{-1}$ beyond the free energy barrier.

\subsection*{E. Molecular Dynamics Simulations}

\subsubsection*{Fullerene C$_{60}$ Dimer}

We performed MD simulations to study the interaction between two fullerene C$_{60}$ molecules solvated by 2398 water molecules, in a simulation box of $3.607 \times 3.607 \times 3.607$~nm$^3$ with periodic boundary conditions. The initial fullerene coordinates and topology are taken from ref. \citenum{Monticelli12}. MD simulations are carried out using GROMACS v2019.4 \cite{Berendsen95, Abraham15} patched with PLUMED 2.5.3 \cite{Tribello14}. We adopted the SPC water model \cite{Berendsen84} and the OPLS-AA force-field \cite{Jorgensen96} for carbon.  Geometry minimization exploited the steepest descent algorithm, stopped when the maximum force was $\leq$ 50 kJ/mol$\cdot$nm. We used the leapfrog algorithm to propagate the equations of motion and the nonbonding interactions were calculated using a PME scheme with a 1.2 nm cutoff for the part in real space. We performed a 100 ps equilibration in an NVT ensemble with a stochastic velocity rescaling scheme \cite{Bussi07} followed by a 100 ps equilibration in an NPT ensemble using the Parrinello-Rahman barostat \cite{Parrinello81} with a time step of 1 fs. We generated MD production trajectories without restrains, with a time step of 1 fs in the NPT ensemble at 298 K and 1 atm. 
The reference free-energy profile of the association/dissociation of the fullerene C$_{60}$ dimer in water as a function of the distance between the centers of mass $(d)$ was computed from five unbiased simulations of 500 ns each, from the population 
histogram: $F(d)=- k_BT \log\rho_\mathrm{eq}(q)$. We estimated the standard deviation of each histogram bin using the five simulations. The resulting profile is consistent with the free-energy profile in ref. \citenum{Monticelli12}. 

The diffusion profile used for comparison was computed from umbrella sampling simulations in which we restrict $d$ with a harmonic potential around a reference value $d_i$ over different windows $i$, 

\begin{equation}
    U(d; d_i) = \frac{c}{2} (d-d_i)^2 
\end{equation}

Then, the diffusion coefficient in the window $i$ is estimated as the ratio between the variance of the variable ($\sigma_{d_i}^2$) and the autocorrelation time of the variable itself ($\tau^{\mathrm{corr}}$):\cite{Hummer05}

\begin{equation}
    D(d_i) = \frac{\sigma_{i}^2}{\tau^{\mathrm{corr}}_i} 
\end{equation}

We setup a window every 0.1 nm in a range of $d$ between 1.0 and 1.8 nm. We performed a 1 ns MD simulation for equilibration and 10 ns MD simulation production in each window with a spring constant $c=1000~\frac{\mathrm{kJ}}{\mathrm{mol}}$. For each window, we split the 10 ns MD simulations in 10 independent blocks of 1 ns each. We estimated the variance of $d$ and the autocorrelation time in each block. Finally, we report the average $D(d_i)$ value and its standard error over the blocks. We note that $D(d)$ does not show large fluctuations with respect to the choice of $c$ (see Figure S2).  

To generate the input trajectories for the construction of Langevin models, we employed aimless-shooting (AS) \cite{Peters06,Peters07} simulations. We obtained a first reactive trajectory by shooting from randomly-picked configurations of the unbiased trajectories with $d$ between 1.2 and 1.3 nm. We then performed AS using a separation $\delta t = 0.1$ ps between successive shooting points, and a total length of 20 ps for each trajectory. We obtained a total of 1110 accepted trajectories relaxing from the transition state region (see Figure \ref{fig:ful-tps}b) with an acceptance rate of 14\%. The script we developed to perform AS with GROMACS is publicly available at \url{https://github.com/physix-repo/aimless-shooting}.

To estimate the dissociation rate we use the reactive flux formalism over 1000 aimless shooting trajectories. 
We define the dissociated state as $d\geq 1.34$ nm and the associated state as $d\leq 1.17$ nm (see Figure \ref{fig:ful-tps}a). The value of the MFPT obtained with reactive flux was validated with the brute force estimate from 100 unbiased MD trajectories launched from the bound state: $6.1 \pm 1.2$~ns versus $6.5 \pm 0.6$~ns, respectively. Finally, we estimated the transmission coefficient as \cite{Jungblut16}

\begin{equation}
    \kappa = \frac{k_{A \rightarrow B}^{RF}}{k_{A \rightarrow B}^{TST}}~,
\end{equation}

where $k_{A \rightarrow B}^{RF}$ is the dissociation rate constant from reactive flux (eq \ref{eq:reactiveflux}), and $k_{A \rightarrow B}^{TST}$ is the dissociation rate constant from transition state theory. The estimated transmission coefficient for the dissociation of the fullerene C$_{60}$ dimer is 0.37, meaning that the number of recrossings before relaxation is small.

\subsubsection*{Fullerene C$_{240}$ Dimer}

We also simulated the interaction between bigger fullerenes, featuring a more stable complex: two C$_{240}$ molecules in water solution. We used the structure of the fullerene molecule from ref. \citenum{Noel14}. We built the topology automatically using the OBGMX web server \cite{Garberoglio12}, solvating the fullerenes with 5375 water molecules in a simulation box of $5.22 \times 5.22 \times 5.22$~nm$^3$, conserving the same parameters for the Lennard-Jones potentials used in the simulation of the C$_{60}$ fullerenes. The parameters of the MD simulations also remain the same as for C$_{60}$.

Given the stability of the bound state, to estimate the reference free energy profile we used well-tempered metadynamics simulations \cite{Barducci08}, adding bias over the distance between the center of mass ($d$) with an initial Gaussian height of 1.0 kJ/mol and  width of 0.01 nm. The bias factor was set to 8, and Gaussians were deposited every 1 ps. Five replicas of 500 ns each were simulated. We estimate the reference free energy profile and its statistical uncertainty from the average and standard deviation of the free energy profiles from all simulations. We obtained the diffusion profiles for comparison from umbrella sampling simulations following the same procedure and with the same parameters used for C$_{60}$ fullerene dimer. We placed a window every 0.1 nm in a range of $d$ from 1.6 to 2.6 nm. 
We performed AS simulations following the same procedure used for C$_{60}$ fullerenes. We obtained 1515 accepted trajectories of 100 ps each, with an acceptance rate of 15\%. 
We calculate the dissociation rate using 1000 trajectories and reactive flux: $9.4 \pm 1.1$~$\mu$s. We define the dissociated state as $d\geq2.01$ nm and the associated state as $d\leq 1.9$ nm. The estimated transmission coefficient is 0.6.

\section*{Results  and Discussion}

In the new approach, the Langevin model is parametrized from unbiased MD trajectories spanning the relevant CV region. Such trajectories can be easily obtained as spontaneous out-of-equilibrium relaxation from initial high free-energy configurations, e.g., the top of a barrier.
Several effective techniques are available to discover transition states and reactive paths at a moderate computational cost \cite{Izrailev99,Bolhuis02,Laio02,Best05,Samanta14,Jung17}: although not trivial, this task is generally much less involved than reconstructing accurate free energy profiles. In the following we use as input  CV-projected MD trajectories $q(t)$ obtained from TPS or by shootings from the barrier top and relaxing into the minima. Note that such trajectories are short, since in rare events the transition path time is typically very fast compared to the waiting time in a minimum, that is, the MFPT.

\begin{figure}[htb!]
\centering
\includegraphics[width=8.4cm]{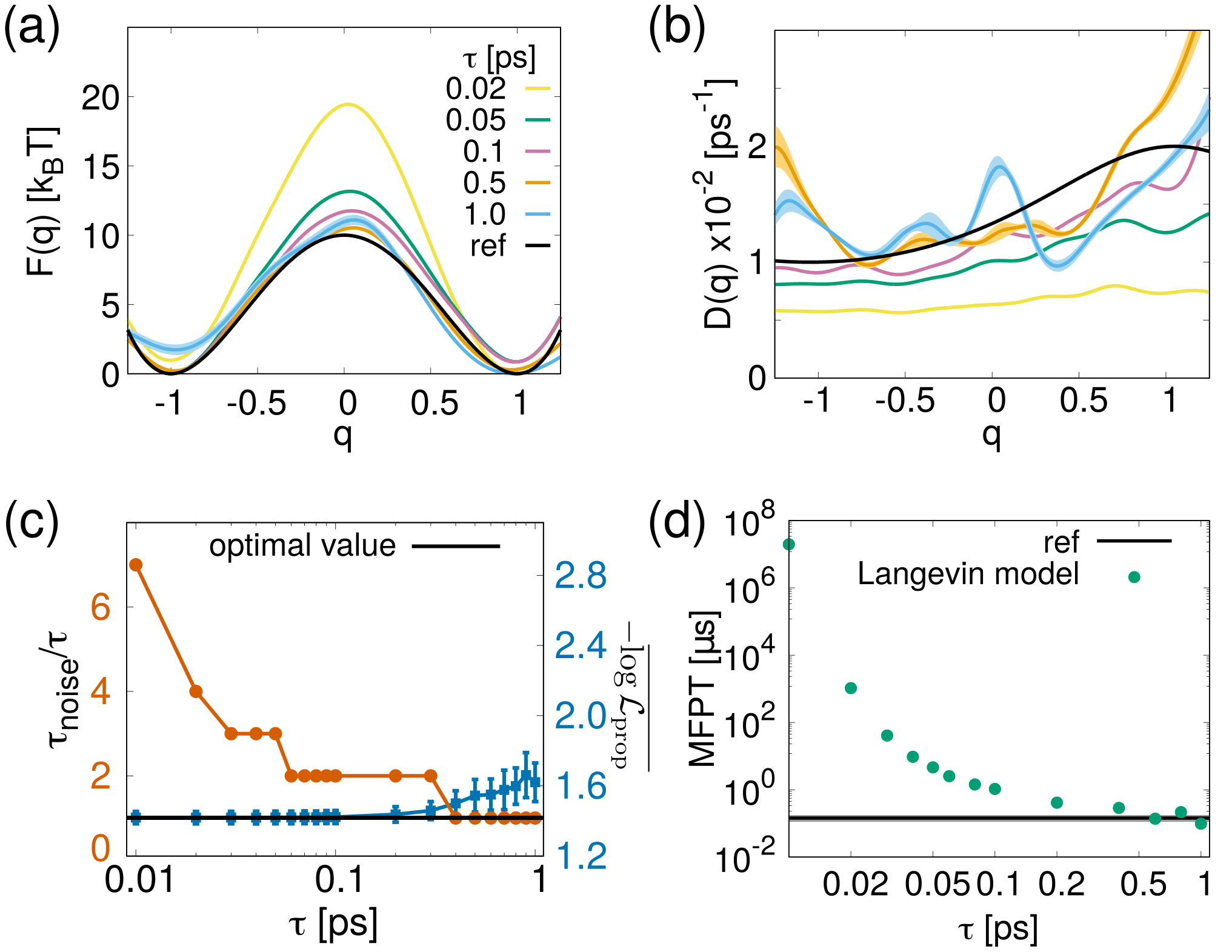}
\caption{Optimal overdamped Langevin models (eq \ref{eq:OLEposdep}) for different resolutions $\tau$ starting from underdamped trajectories on a double-well free-energy landscape.
(a) $F(q)$ and (b) $D(q)$ profiles (exact ones are shown with black lines);
vertical bars correspond to standard error over 10 independent stochastic optimizations.
(c) Autocorrelation time of the effective noise $\tau_\mathrm{noise}$ (eq \ref{eq:effnoise-m}) normalized by $\tau$ (i.e., number of correlated steps), and average likelihood quantifying the accuracy of the propagator (eq \ref{eq:Lprop}). The black line represents the ideal values.
Vertical bars are the standard deviation over all trajectory steps $\Delta q$.
(d) MFPTs for the optimal overdamped Langevin models shown in panels a and b, computed with eq \ref{eq:mfpt-Smoluchowski}: the reference value in black is computed with reactive flux on the exact profiles in black in panels a and b. Vertical bars are the standard error over 10 independent Langevin models (hardly visible at this scale).
}\label{fig:dw-res}
\end{figure}

As an initial benchmark we verify whether our method yields Langevin models accurately reproducing the correct $F(q)$ and $D(q)$ starting from underdamped Langevin trajectories. The input data are composed by 100 short underdamped Langevin trajectories $q(t)$, each one 20 ps long, relaxing from the barrier top of a double-well landscape (see Supplementary Figure S3). The exact free energy and diffusion profiles are shown in black in Figure \ref{fig:dw-res}.

Since the initial trajectories are generated with an underdamped integrator, building the overdamped model entails projecting out the momentum $p$ conjugated to the CV position $q$. As a result, a $q$-based Langevin model is expected to display memory (time correlation) in the friction and noise for $\tau$ comparable to the integration time step $\Delta t=10^{-4}$ ps.
Adopting coarse-enough $\tau$ resolutions, on the other hand, should lead to memory-less behavior, well reproduced by an overdamped model.

To test this hypothesis, in Figure \ref{fig:dw-res} we generate Langevin models using $0.01$ ps $\leq \tau \leq 1$ ps: for the smallest $\tau$ value, the model strongly overestimates the barrier and underestimates the diffusion coefficient. As $\tau$ increases the reconstructed model becomes more accurate (see Figure \ref{fig:dw-res}) until at large values of $\tau$ the error increases again. 
We found that there is a relatively narrow range of $\tau$ values for which the Langevin overdamped model is accurate: 
to identify the optimal range in real-world applications we propose two complementary tests that allow identifying the upper and lower boundaries of $\tau$ bracketing the high-quality Langevin models.

The lower acceptable $\tau$ is the smallest time displaying no memory effects, measured through the autocorrelation of the effective Langevin noise, that is, the noise extracted a posteriori by analyzing the original trajectory with the optimal Langevin model, eq \ref{eq:effnoise-m} (see Figure S1 for details).
The upper limit on $\tau$ corresponds to the onset of a significant deviation between the approximate and the exact propagators (the latter estimated from accurate numerical integration, see methods section C and eq \ref{eq:Lprop}). Both diagnostics are reported in Figure \ref{fig:dw-res}c, and together they predict that the Langevin model should attain the best accuracy for $\tau$ values close to 0.5 ps: large enough to avoid memory, but small enough for the short-time propagator to remain reliable.

Overall, when selecting $\tau$ based on the proposed diagnostics, models converge smoothly and rapidly ($<10^6$ Monte Carlo steps) to a good approximation of the exact results: as few as 100 reference trajectories are sufficient to reconstruct the free energy profile to within 1 $k_BT$ and $D(q)$ to within 15\% error (we remark that fluctuations in $D(q)$ appear converged with respect to the different parameters used in the reconstruction of the models and irrespective of the input trajectories). The same conclusions can be derived regardless of the shape of the diffusion and free energy profiles (see Figure S4). We remark that, when the input trajectories are generated with an overdamped integrator, the reconstructed profiles deviate less than 1 $k_BT$ and 10\%, respectively, from the exact $F(q)$ and $D(q)$, regardless of $\tau$, unless when it is too large for our approximate propagator (see Figure S5).

Besides estimating free energy and diffusion profiles, a key objective of the new method is the accurate prediction of kinetic rates at a moderate computational cost. 
Figure \ref{fig:dw-res}d shows the MFPTs obtained with eq \ref{eq:mfpt-Smoluchowski} using the optimal Langevin models.
Too small values of $\tau$ lead to strongly overestimated MFPT, as expected from the overestimated barriers, while the rate becomes accurate for $\tau$ values within the predicted optimal window (Figure \ref{fig:dw-res}c).
On the other hand, for $\tau$ larger than optimal, MFPTs in the same order of magnitude as the reference value are still obtained.

We note that, alternatively, we can compute the MFPT using the reactive flux technique \cite{Chandler78,Hanggi90}, free from transition state approximations.~
This traditionally requires two distinct sets of expensive simulations, one to compute the free energy barrier and another to compute a correlation function accounting for recrossings (see methods section D and Table S1). However, this technique becomes inexpensive if combined with our approach, since the free energy barrier is directly obtained from likelihood optimization on short trajectories, while recrossing statistics can be accurately estimated from inexpensive synthetic Langevin trajectories.
We remark that the rates computed from  eq \ref{eq:mfpt-Smoluchowski}  are similar to those of reactive flux (see Table S2).

\begin{figure}[htb!]
\centering
\includegraphics[width=8.4cm]{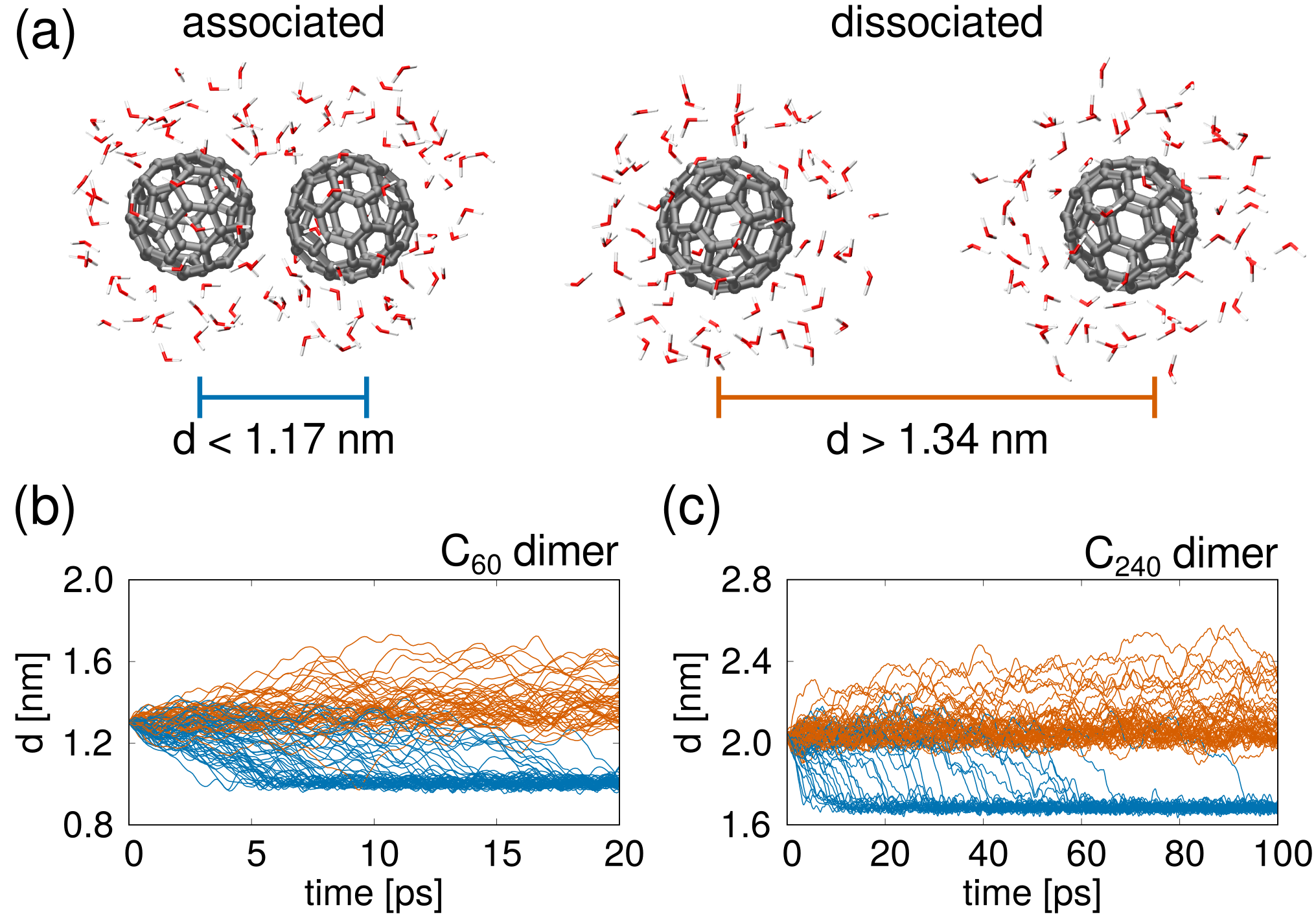}
\caption{
(a) Characteristic atomic configurations for the C$_{60}$ dimer in bulk water solution; only waters distant less than 5 \AA~from carbon atoms are shown. The associated and dissociated states are defined in terms of the distance $d$ between the centers of mass of the fullerenes.
(b) 100 MD trajectories (from aimless shooting) starting from the transition state ensemble and featuring the association (blue) or dissociation (orange) of the C$_{60}$ dimer, and (c) of the C$_{240}$ dimer, respectively. 
}\label{fig:ful-tps}
\end{figure}

\begin{figure}[htb!]
\centering
\includegraphics[width=0.9\textwidth]{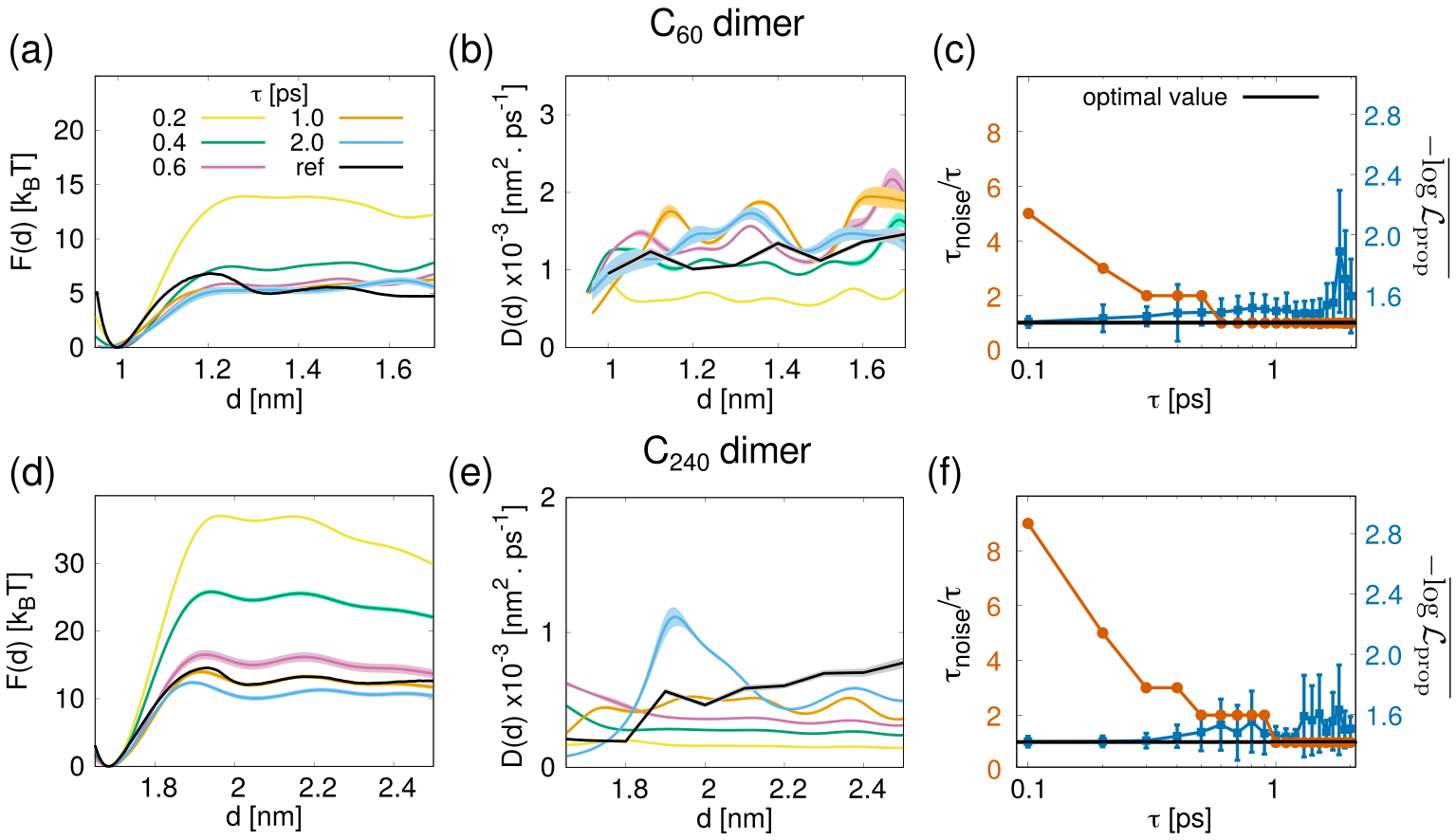}
\caption{
Optimal Langevin models for the interaction of C$_{60}$ and C$_{240}$ fullerene dimers in explicit water, for different values of the time resolution $\tau$. The input data set consists of 100 MD trajectories (aimless shooting) relaxing from the transition state ensemble in Figure \ref{fig:ful-tps}. 
(a, d) Free-energy profiles $F(q)$, compared to the reference ones in black:
vertical bars correspond to standard deviations over 10 independent stochastic optimizations.
(b, e) Diffusion profiles $D(q)$, compared to the umbrella sampling ones in black:
vertical bars correspond to standard deviations over 10 independent stochastic optimizations.
(c, f) Autocorrelation time of the effective noise (eq \ref{eq:effnoise-m}) normalized by $\tau$ (i.e., number of correlated steps), and average likelihood quantifying the accuracy of the propagator (eq \ref{eq:Lprop}). The black line represents the ideal value for both quantities; vertical bars refer to the standard deviation over all trajectory steps $\Delta q$.
}\label{fig:ful-res}
\end{figure}

The double-well potentials in the previous section allowed us to benchmark the scope and limitations of the method while having full control over the system. To assess the method in realistic complex systems, we modeled the dynamics of two fullerene molecules in explicit water, considering two different sizes C$_{60}$ and C$_{240}$ (Figure \ref{fig:ful-tps}). The free-energy landscape as a function of the distance $d$ between fullerenes' centers of mass features a minimum for the bound complex (with different depth for the two sizes) and a relatively flat region for the dissociated state, see black lines in Figure \ref{fig:ful-res}. The resulting rare dissociation events and diffusion-controlled association events  \cite{Banerjee13,Li05a,Chiu10,Monticelli12} are analogous to other processes such as protein-protein interaction. 
We obtained reference free-energy profiles using 2.5~$\mu$s-long brute-force MD trajectories for the C$_{60}$ dimer, while we employed well-tempered metadynamics \cite{Barducci08} for the C$_{240}$ dimer due to its very low dissociation rate. The diffusion profiles are obtained from sets of 10 ns umbrella sampling simulations every 0.1 nm in distance between the center of mass (see methods section E for details).

We maximized the likelihood of Langevin models starting from short aimless shooting \cite{Peters06} TPS trajectories (20/100 ps each, cumulative duration 2/10 ns for C$_{60}$/C$_{240}$, see Figure \ref{fig:ful-tps}). An example of the convergence of $-\log \mathcal{L}(\theta)$ as a function of the number of iterations is shown in Figure S6.
We adopted a range of time resolutions $\tau$ between 0.1 and 2 ps: for fast time scales, memory is expected to arise from the projection of the solvated fullerenes many-body dynamics on a single CV.
Sizable memory effects are evident in the effective noise (eq \ref{eq:effnoise-m}) for $\tau<0.5$ ps (see also Figure S1), severely affecting the accuracy of $F(d)$. For larger $\tau$ values, similarly to the case of the analytic double-well potential, the Langevin models become increasingly accurate: memory decays after about 0.5 ps for the smaller fullerenes and after about 1.0 ps for the bigger ones. As for the diffusion profile, all models show $D(d)$ in the same order of magnitude, in good agreement with the diffusion coefficients estimated from umbrella sampling \cite{Hummer05} (see methods section E). 
We remark that generating 100 short TPS trajectories for Langevin optimization can be computationally faster than generating a sufficient amount of umbrella sampling trajectories to estimate the diffusion profile.

On the other hand, the quality of the approximate short-time propagator eq \ref{eq:prop2} decreases as $\tau$ increases.
Accurate results are clearly identified in the range of $\tau$ that minimizes both non-Markovian behavior and error propagation (see Figure \ref{fig:ful-res}). 
Interestingly, optimizing the models based on 1000 reference trajectories instead of 100 yields minimal improvements (see Supplementary Figure S7).

\begin{figure}[htb!]
\centering
\includegraphics[width=8.4cm]{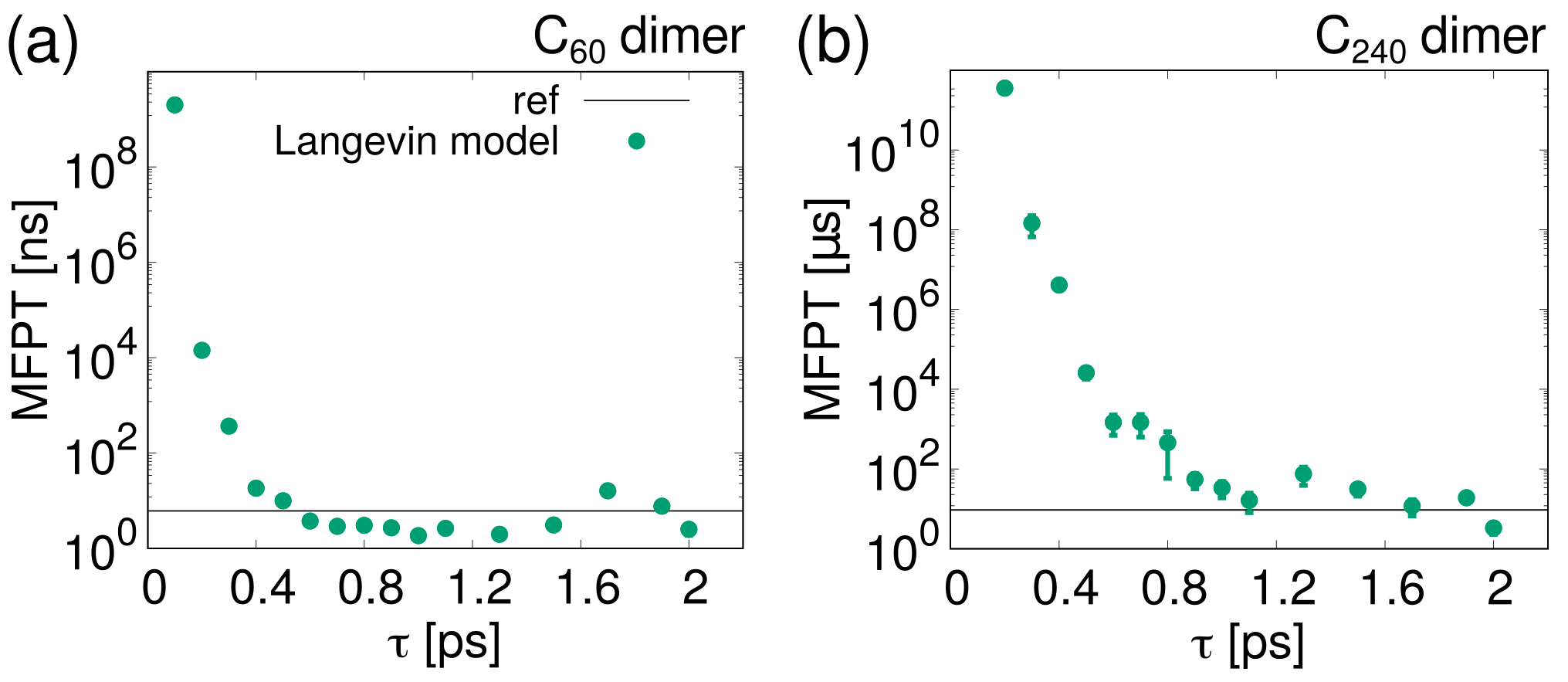}
\caption{
MFPTs computed from the optimal Langevin models generated with different time resolutions $\tau$, for the dissociation of fullerene (a) C$_{60}$, and (b) C$_{240}$ dimer, in water solution.
MFPTs are estimated from eq \ref{eq:mfpt-Smoluchowski} (reference value in black). Vertical bars are the standard error over 10 independent Langevin models.
}\label{fig:ful-rates}
\end{figure}

Finally, we assessed the accuracy of kinetic predictions given by the models. The dissociation MFPT is on the nanosecond and microsecond scales, respectively, for C$_{60}$ and C$_{240}$ dimers; that is, 2 and 5 orders of magnitude slower than the out-of-equilibrium MD trajectories used for training. 
For each value of $\tau$, we computed MFPTs using eq \ref{eq:mfpt-Smoluchowski} on the optimal models, yielding Figure \ref{fig:ful-rates}. 
We emphasize that the latter approach allows computation of the MFPT without generating any additional expensive MD trajectories.

For too small values of $\tau$, the barrier is overestimated (see Figure \ref{fig:ful-res}), leading to overestimation of the MFPTs by several orders of magnitude. 
For the C$_{60}$ dimer the MFPT becomes accurate (compared to the brute force result) for $\tau > 0.5$~ps, while for the C$_{240}$ dimer this level of accuracy requires $\tau > 1$~ps: once again, these are the optimal time resolutions predicted by the diagnostics in Figure \ref{fig:ful-res}. 
Taken together, all these results demonstrate that the approximations inherent in the Langevin models are under control, leading to accurate predictions about the thermodynamics and kinetics of complex systems.

\section*{Concluding Remarks}

The calculation of free energy landscapes and kinetic rates is key tasks of computer simulations of complex systems. Even though these two tasks are usually tackled using different {\sl ad hoc} techniques,\cite{Camilloni18}
the main contribution of the present work consists in demonstrating that they can be {\sl simultaneously} achieved in a conceptually simple way, by estimating Langevin models starting from a relatively inexpensive set of about one hundred TPS-like trajectories, regardless of the barrier height.
Likelihood maximization -- an efficient parameter estimation technique -- is combined with a double test bracketing the time resolution $\tau$, granting control over the accuracy of the model. 
The choice of a suitable $\tau$ is especially important for accurately estimating the free energy, because the rate has an exponential dependence on the barrier while only having a linear dependence on $D$.

The resulting Markovian Langevin equations reproduce well the quantitative thermodynamic and dynamic properties of the original many-body system, including accurate kinetic rates, despite a gap of many orders of magnitude with respect to the short MD trajectories used for training. Moreover, the short time transition probability $p(q',t+\tau | q,t)$ predicted by the optimal Langevin model (eq \ref{eq:prop}) is in good agreement with a distribution collected from short MD shootings of length $\tau$ (see example in Figure S8). Incidentally, we note that our approach is not restricted to transition path sampling: in principle, any set of unbiased trajectories spanning the regions of interest could be used to train the Langevin model (see example in Figure S9). 

We also remark that equilibrium properties are systematically recovered from out-of-equilibrium data: standard transition path sampling trajectories are the golden standard for the study of transformation mechanisms, however, since 
such data set is a small subset of all possible (reactive and nonreactive) pathways, lacking Boltzmann distribution of the configurations, it cannot be used for the {\sl direct} estimate of equilibrium histograms (i.e., free energies) and rate matrices (e.g., in Markov state models) by simple averaging \cite{Bolhuis10,Bolhuis21}. Here we show that the contrary is true, provided bare transition paths are employed to train a suitable stochastic model (see also refs. \citenum{Hummer03,Sriraman05, Hummer05,Zhang11,Innerbichler18,Brotzakis19} for related or alternative ideas). Note that Langevin equations of motion, compared to other machine-learning tools, retain a direct physical interpretation, with the separation between a systematic average force and friction/noise effects describing the projected-out degrees of freedom (commonly referred to as ``the bath''). 

In future applications of the new approach, two main issues have to be taken into consideration. First, different kinds of Langevin equation (generalized, standard or overdamped) can be necessary to faithfully reproduce the projected dynamics of a complex system, depending on the physical process, the choice of the CV, and the observational time scale. 
For chemical reactions in water the memory time scale could be comparable to the transition path time, requiring a non-Markovian equation, whereas for crystal nucleation or protein folding Markovian models are customarily invoked.
In the present work, we demonstrated the approach using overdamped Langevin equations: in principle, this can be generalized to other stochastic models by replacing the propagator expression (eq \ref{eq:prop2}) with a different suitable approximation, based on ideas in refs. \citenum{Drozdov97,Vroylandt22}.

A second issue concerns the effect of projecting the dynamics on suboptimal CVs, rather than on the ideal reaction coordinate, commonly identified with the committor function \cite{Lu14,Banushkina16,Peters16}. 
In principle, the CV definition can be systematically optimized with an iterative scheme, in which a single initial set of MD trajectories can be employed to build optimal Langevin models for progressively improving CVs. A related idea was recently proposed in the case of discrete Markov state models \cite{Tiwary16,Chen18}.
Therefore, our new approach represents a starting point both for the systematic modeling of MD data through physically-motivated stochastic equations as well as for machine-learning approaches to reaction coordinate optimization.

\section*{Acknowledgement}

We gratefully acknowledge very insightful discussions with Christoph Dellago, Gerhard Hummer, Gerhard Stock, A. Marco Saitta, Rodolphe Vuilleumier, Hadrien Vroylandt, Riccardo Ferrando, Alessandro Barducci, and Andrea Pérez-Villa. F.P. acknowledges the Erwin Schrödinger International Institute for Mathematics and Physics at the Universität Wien, for a stay in April 2019. Calculations were performed on the GENCI-IDRIS French national supercomputing facility, under Grant Nos. A0070811069, A0090811069, and A0110811069.

\section*{Supporting information}

The Supporting Information is available free of charge at \url{https://pubs.acs.org/doi/10.1021/acs.jctc.2c00324}.
Detailed results for different shapes of the double-well benchmark systems, autocorrelation functions of the effective noise, convergence tests, and kinetic rate estimates using reactive flux

The authors declare no competing financial interest.

\bibliography{ms}

\newpage

\section*{Supporting Information}

\setcounter{table}{0}
\renewcommand{\thetable}{S\arabic{table}}%
\setcounter{figure}{0}
\renewcommand{\thefigure}{S\arabic{figure}}%

\section*{Supplementary Figures}

\begin{figure*}[htb!]
\centering
\includegraphics[width=\textwidth]{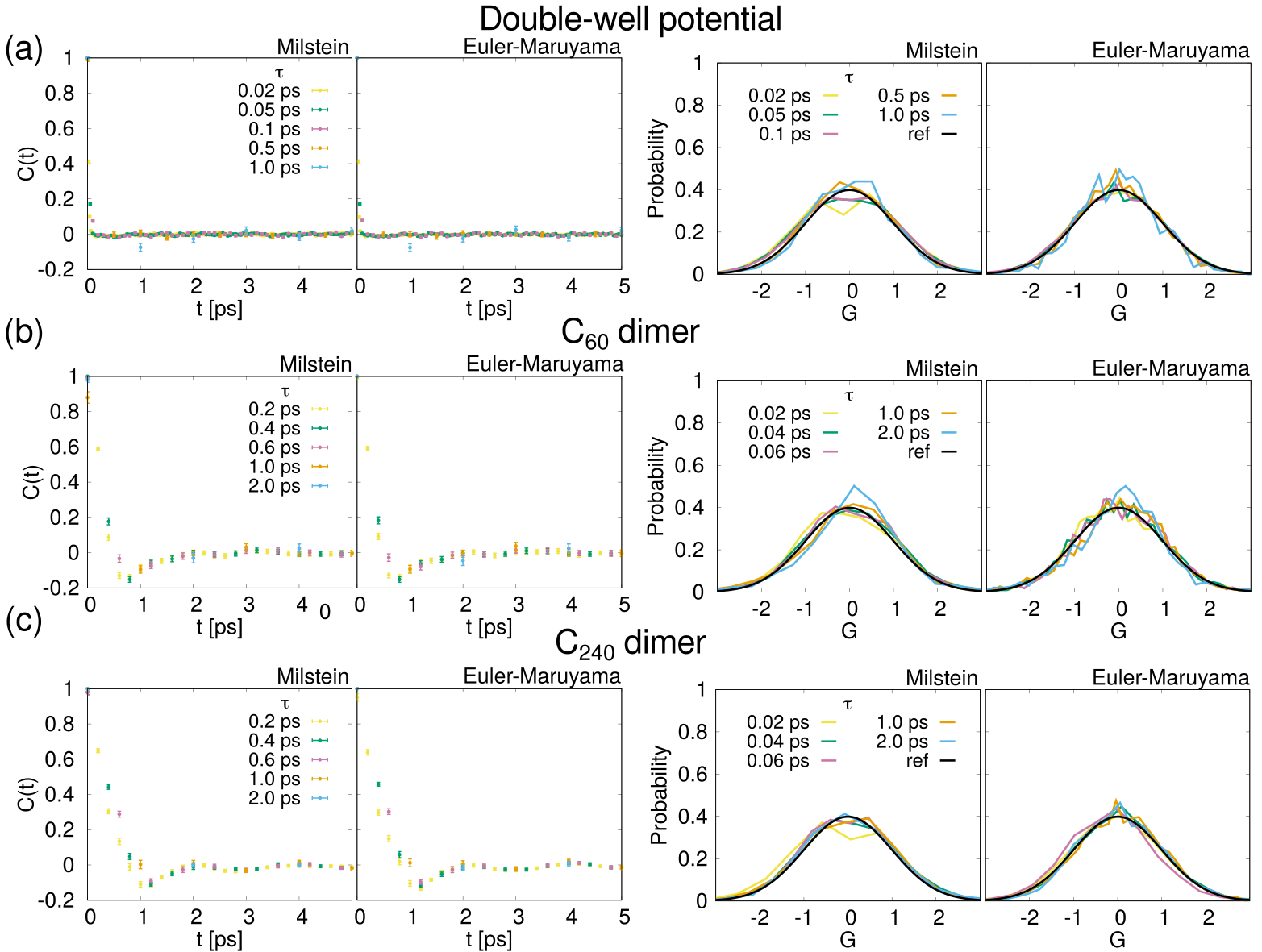}
\caption{Autocorrelation function $C(t)$ (left panels) and distribution of the effective noise $G$ (right panels) for optimal Langevin models of different systems. The expression used to estimate the noise is found inverting the Euler-Maruyama, or the Milstein integrator, see the Methods section C, in the Main Text. (a) Symmetric double-well potential (non-overdamped input trajectories) with 10~$k_BT$ barrier and diffusion coefficient $0.01 \leq D(q) \leq 0.02$~ps$^{-1}$ (see Fig. S3 and Fig. 2 in the Main Text).
(b) C$_{60}$ fullerene dimer in solution.
(c) C$_{240}$ fullerene dimer in solution.
The different time resolutions $\tau$ used to optimize the overdamped Langevin models are shown in colors. 
The expected zero-mean, unit-variance Gaussian distribution of the noise is shown in black lines on the right panels.
Error bars correspond to standard error over all the  trajectories data-points.}\label{fig:act-noise}
\end{figure*}

\begin{figure*}[htb!]
\centering
\includegraphics[width=0.7\textwidth]{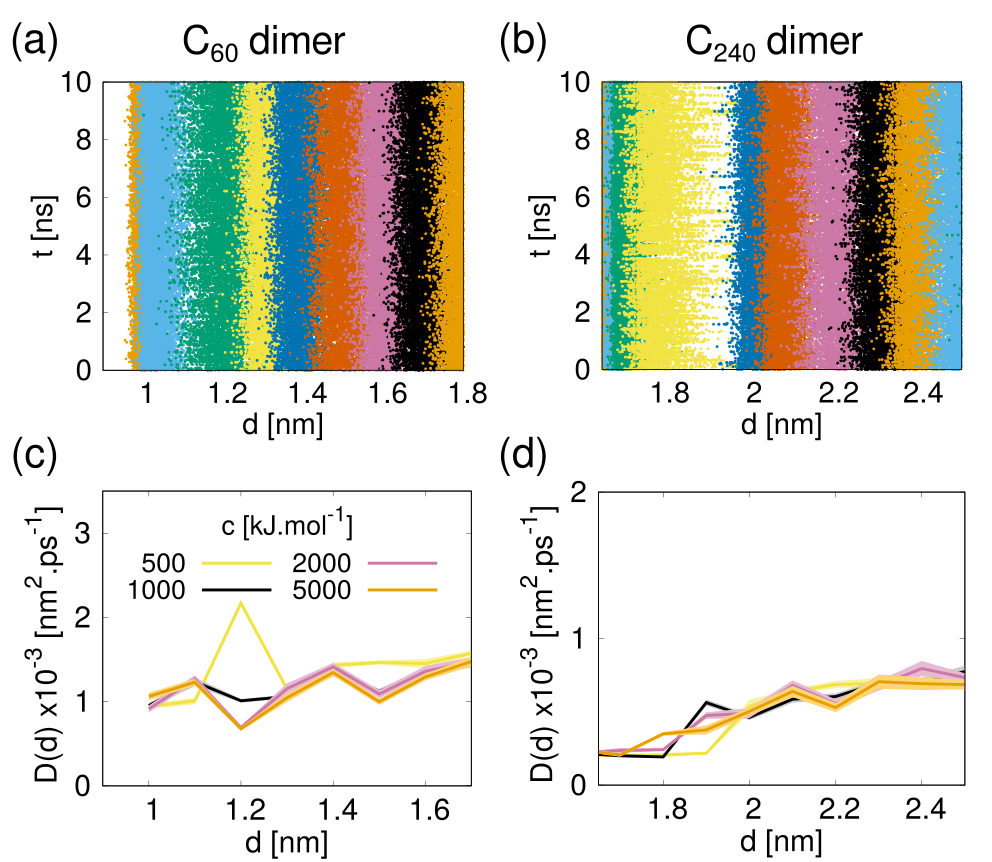}
\caption{Diffusion profiles from umbrella sampling. Trajectory on each window for
(a) C$_{60}$ fullerene dimer and (b) C$_{240}$ fullerene dimer using a spring constant $c=1000~\mathrm{kJ \cdot mol^{-1}}$. Note that for the calculation of the diffusion coefficient we do not required converged umbrella sampling calculations. Diffusion profiles $D(d)$ for different values of the spring constant $c$ for (c) C$_{60}$ fullerene dimer and (d) C$_{240}$ fullerene dimer.
}\label{fig:us}
\end{figure*}

\begin{figure*}[htb!]
\centering
\includegraphics[width=0.4\textwidth]{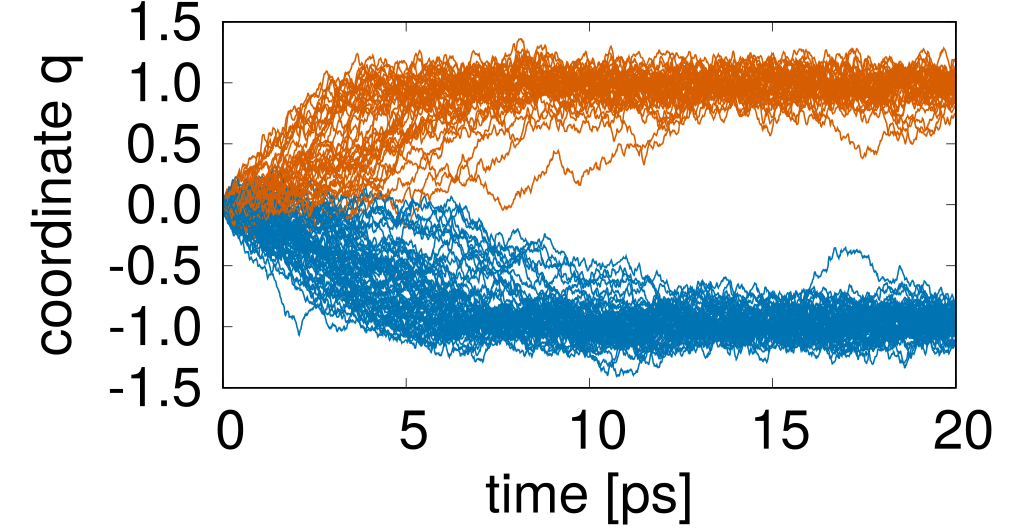}
\caption{100 input trajectories generated with a non-overdamped Langevin integrator, using the standard Langevin equation: $\dot p = -\frac{\partial F(q)}{\partial q} -\gamma p + \sqrt{2k_BTm\gamma}\,\eta(t)$. The benchmark system is a symmetric double-well free-energy landscape with 10 $k_BT$ barrier ($m=1~k_BT~\cdot$ ps$^{-2}$). 53\% of the trajectories ended up in the right basin and 47\% in the left basin.}\label{fig:trajs-nover}
\end{figure*}

\begin{figure*}[htb!]
\centering
\includegraphics[width=\textwidth]{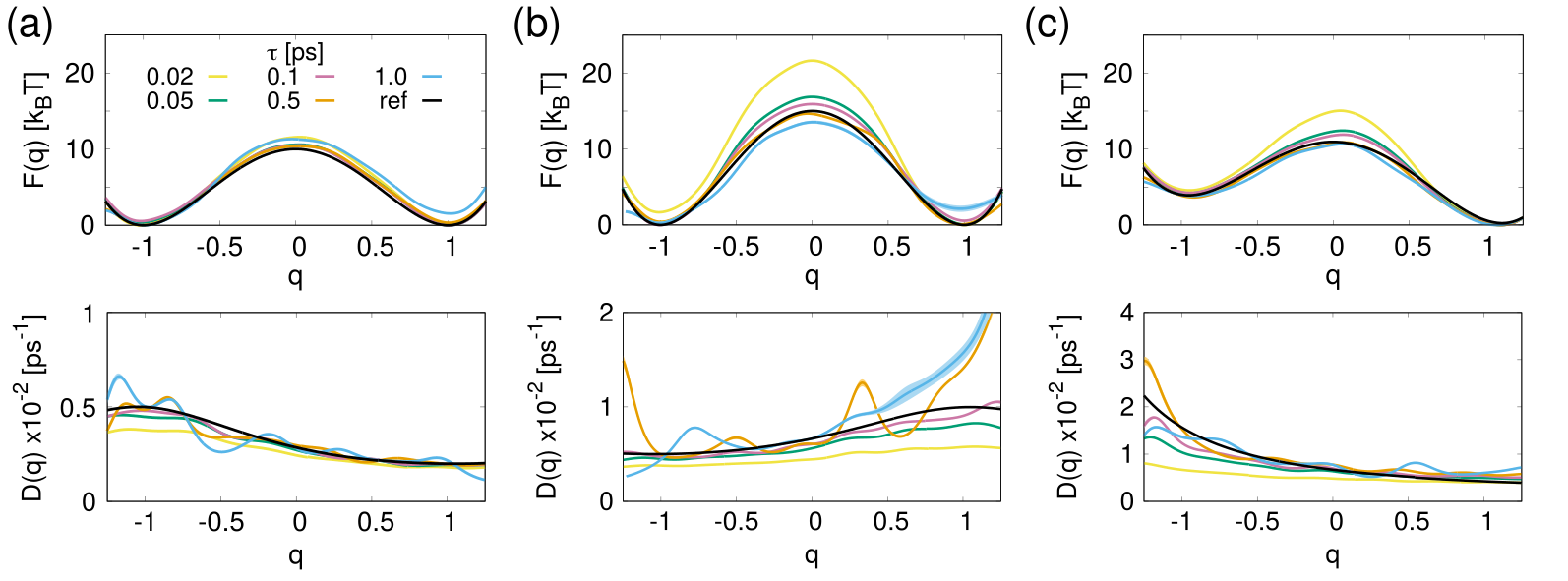}
\caption{Optimal overdamped Langevin models, each one trained on 100 non-overdamped Langevin trajectories of 50 ps relaxing from the barrier top.
(a) Symmetric double-well potential with 10~$k_BT$ barrier and diffusion coefficient $0.002 \leq D(q) \leq 0.005$~ps$^{-1}$.
(b) Symmetric double-well potential with 15~$k_BT$ barrier and diffusion coefficient $0.005 \leq D(q) \leq 0.01$~ps$^{-1}$.
(c) Asymmetric double-well potential with 10~$k_BT$ barrier and diffusion coefficient $0.005 \leq D(q) \leq 0.03$~ps$^{-1}$.
The different time resolutions $\tau$ employed to optimize the models are shown in colors. 
The exact profiles are shown with black lines.
The thickness of the lines correspond to standard error over 10 independent stochastic optimization runs.
}\label{fig:res-dw-nover}
\end{figure*}

\begin{figure*}[htb!]
\centering
\includegraphics[width=\textwidth]{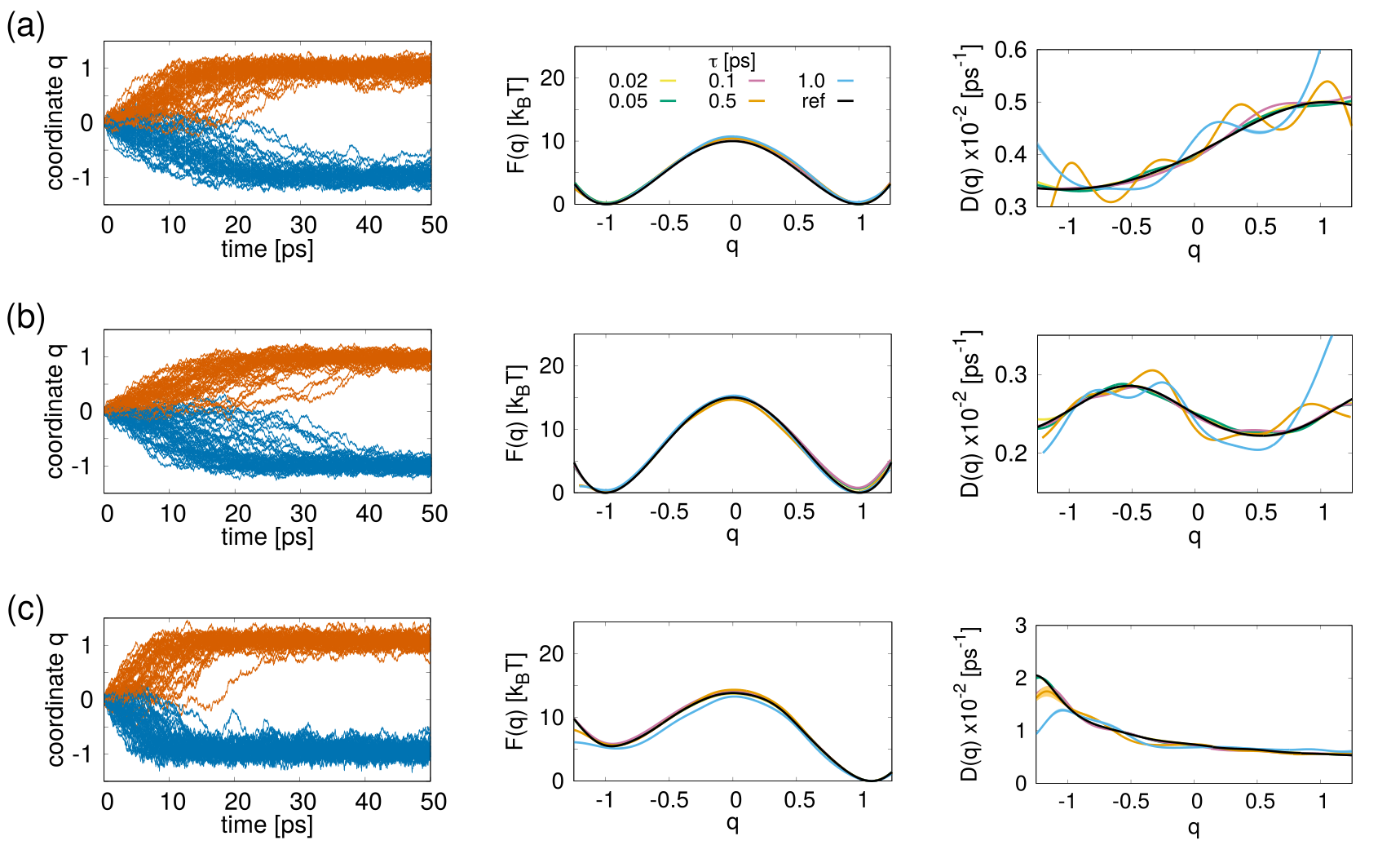}
\caption{Optimal overdamped Langevin models trained on overdamped Langevin trajectories.
(a) Symmetric double-well potential with 10~$k_BT$ barrier and diffusion coefficient $0.003 \leq D(q) \leq 0.005$~ps$^{-1}$.
(b) Symmetric double-well potential with 15~$k_BT$ barrier and diffusion coefficient $0.0022 \leq D(q) \leq 0.0028$~ps$^{-1}$.
(c) Asymmetric double-well potential with 13~$k_BT$ barrier and diffusion coefficient $0.005 \leq D(q) \leq 0.025$~ps$^{-1}$.
Left: 100 input trajectories of 50 ps obtained with an overdamped integrator, Eq. 8 in the Main Text, relaxing from the barrier top.
Center: optimal free energy profiles.
Right: optimal diffusion profiles.
The different time resolutions $\tau$ employed to optimize the models are shown in colors. 
The exact profiles are shown with black lines.
The thickness of the lines correspond to standard error over 10 independent stochastic optimization runs.}\label{fig:res-dw-over}
\end{figure*}

\begin{figure*}[htb!]
\centering
\includegraphics[width=0.5\textwidth]{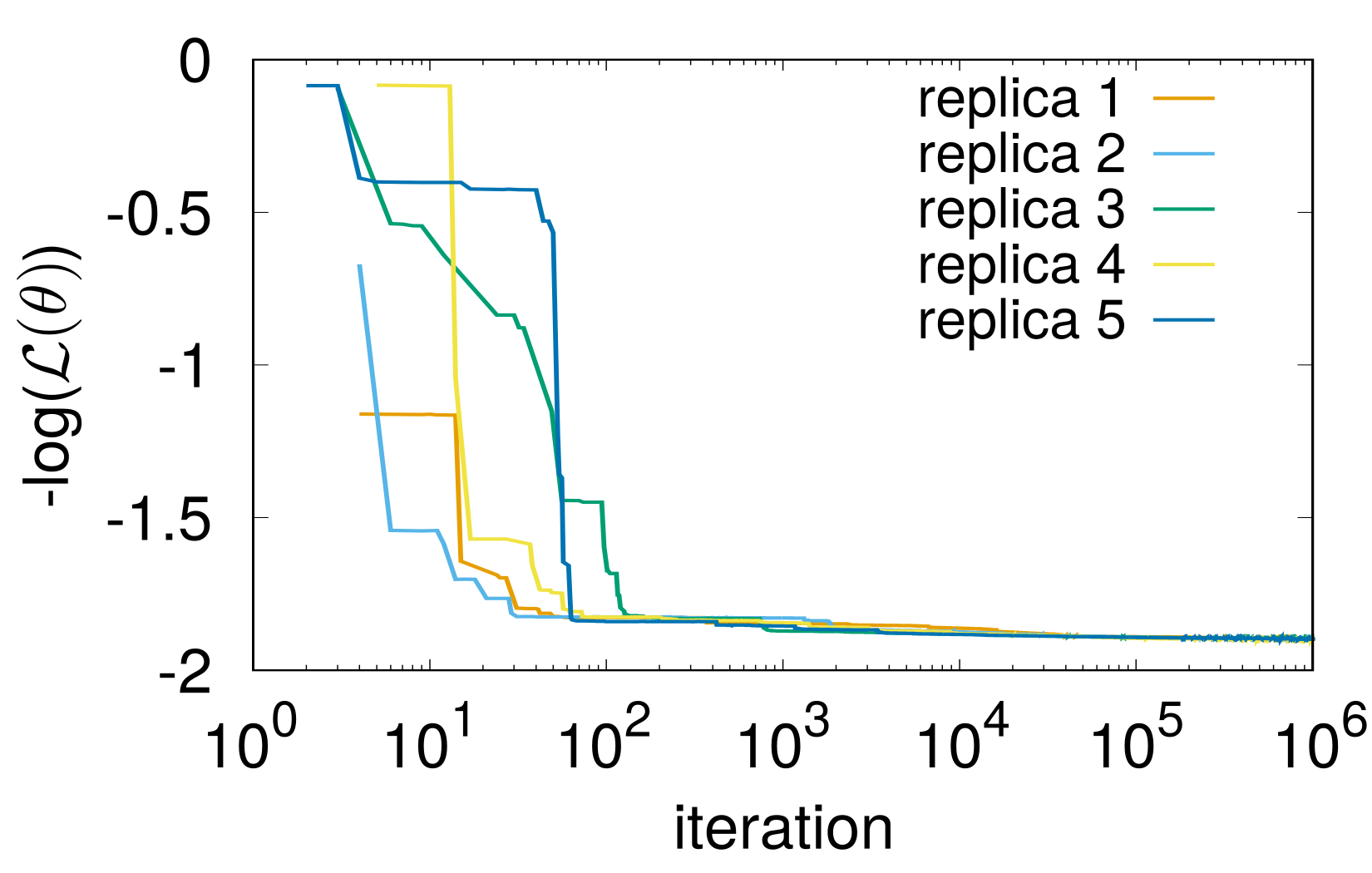}
\caption{Example of the convergence of $-\log \mathcal{L}(\theta)$ as a function of the number of iterations for the C$_{60}$ fullerene dimer in solution.
Colors represent different independent optimizations. A similar behavior with a smooth convergence is observed in all the Langevin models presented in this work.}\label{fig:optimization}
\end{figure*}

\begin{figure*}[htb!]
\centering
\includegraphics[width=0.7\textwidth]{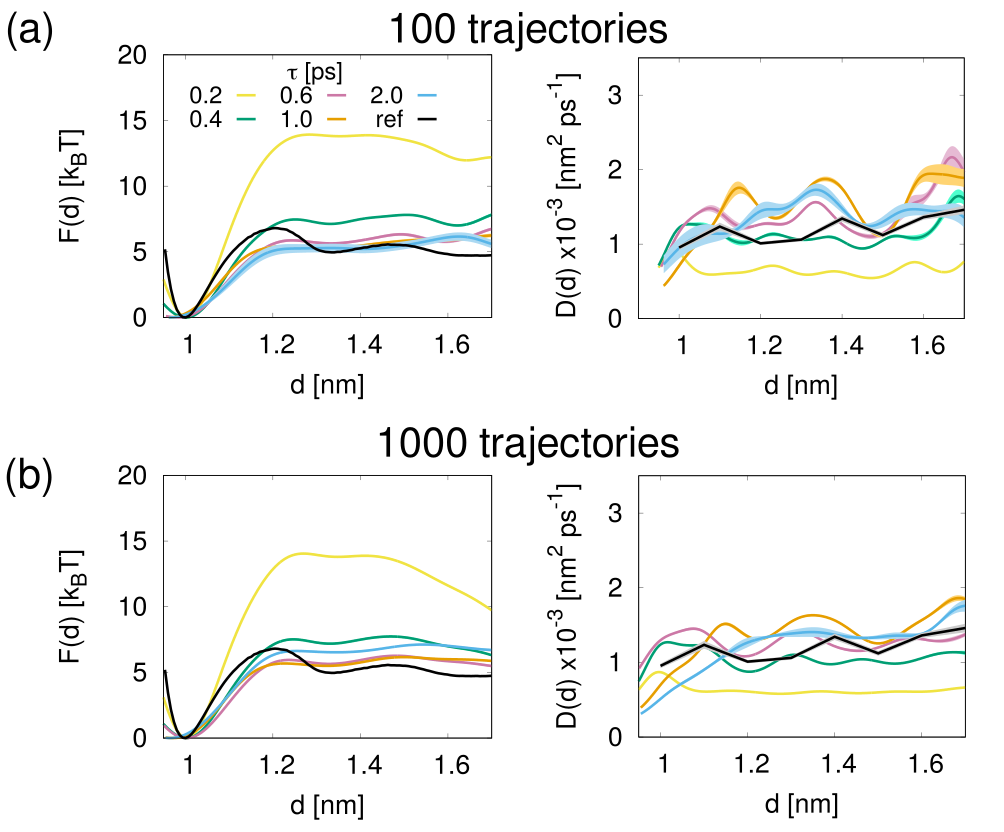}
\caption{Effect of the number of input trajectories. Optimal free energy and diffusion profiles reconstructed from (a) 100 and (b) 1000 initial MD trajectories of 20 ps for the C$_{60}$ fullerene dimer.
The different time resolutions $\tau$ employed to optimize the models are shown in colors. 
The reference free energy profile (computed from brute force MD) and diffusion profile (computed from umbrella sampling) are shown with black lines.
The thickness of the lines correspond to standard error over 10 independent optimization runs.}\label{fig:res-1000trajs}
\end{figure*}

\begin{figure*}[htb!]
\centering
\includegraphics[width=0.5\textwidth]{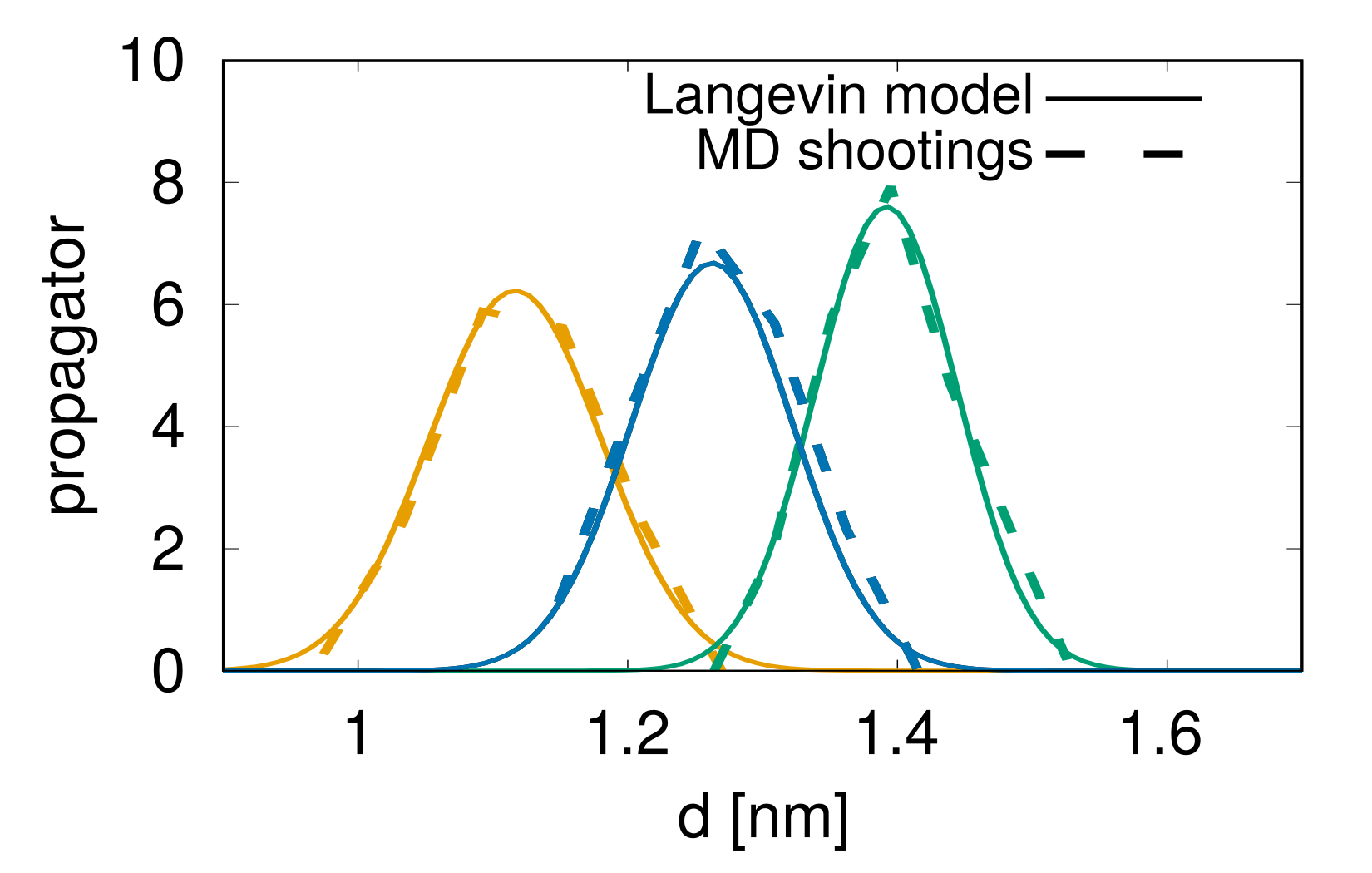}
\caption{Propagator test. Comparison between $p(q',t+\tau | q,t)$ predicted by the optimal Langevin model and a distribution of MD shootings trajectories for the C$_{60}$ fullerene dimer. The time resolution was set to $\tau = 1$~ps. The optimized free energy and diffusion profiles (Fig. 4.a. Main Text, orange profiles) were used to estimate the short time transition probability (eq. 4. in the Main Text) predicted by the Langevin model (solid lines). The distribution from MD correspond to the histograms of the displacement after 1 ps from sets of 500 shooting trajectories (dashed lines). We use 5 shooting points extracted from a long unbiased trajectory for each location: before the barrier (orange), at the barrier top (blue), and after the barrier (green).}\label{fig:test-prop-rev}
\end{figure*}

\begin{figure*}[htb!]
\centering
\includegraphics[width=\textwidth]{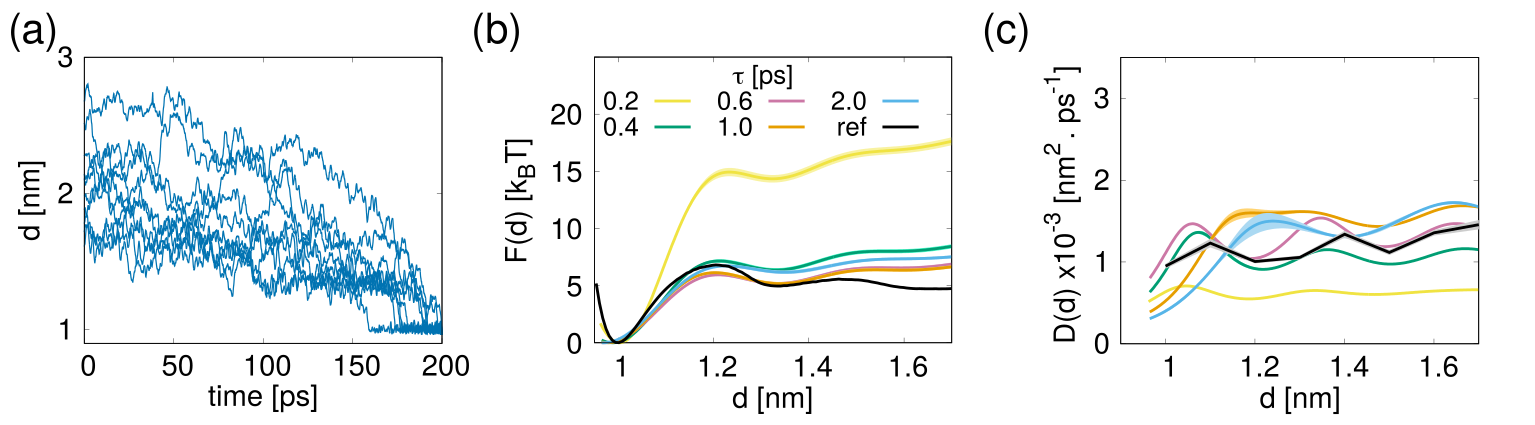}
\caption{Langevin models from association trajectories. (a) Examples of 100 association trajectories used to optimized the (b) free energy and (c) diffusion profiles from 200 ps MD trajectories of the C$_{60}$ fullerene dimer.
The different time resolutions $\tau$ employed to optimize the models are shown in colors. 
The reference free energy profile (computed from brute force MD) and diffusion profile (computed from umbrella sampling) are shown with black lines.
The thickness of the lines correspond to standard error over 10 independent optimization runs.}\label{fig:res-association}
\end{figure*}

\clearpage

\section*{Supplementary Tables}

\begin{table}[htb!]
    \centering
    \begin{tabular}{|c|c|c|}
    \hline
        Method                         & N. trajs & MFPT (ns)           \\ \hline
        Brute force                    & 100      & 18.2 $\pm$ 0.1      \\ \hline
        \multirow{9}{*}{Reactive flux} & 1000     & 90 $\pm$ 69       \\ \cline{2-3} 
                                       & 2000     & 25 $\pm$ 4       \\ \cline{2-3} 
                                       & 5000     & 27 $\pm$ 2      \\ \cline{2-3} 
                                       & 8000    & 20.8 $\pm$ 0.3      \\ \cline{2-3} 
                                       & 10000    & 19.1 $\pm$ 0.1      \\ \cline{2-3} 
                                       & 12000    & 19.07 $\pm$ 0.09      \\ \cline{2-3} 
                                       & 15000    & 18.24 $\pm$ 0.05      \\ \cline{2-3} 
                                       & 18000    & 18.17 $\pm$ 0.04      \\ \cline{2-3}
                                       & 20000    & 18.20 $\pm$ 0.07      \\
    \hline
    \end{tabular}
    \caption{Brute force vs reactive flux estimate of the MFPT. We used as a benchmark system a double-well potential with a 7 $k_BT$ barrier and $50 \leq \gamma(q) \leq 150$ ps. Brute force estimate is obtained as the average escape time of 100 trajectories (at least 50 ns-long) starting from the minimum. Reactive flux estimate is obtained with Eq. 12 in the Main Text, and different number of trajectories. Each trajectory is 40 ps-long. We note that at least 20000 trajectories are needed to converge the reactive flux estimate. We highlight that in practical applications, the need of large amount of trajectories to converge the correlation function of Eq. 12 in the Main Text restricts the use of reactive flux.}
    \label{tab:bf-vs-rf}
\end{table}

\begin{table}[htb!]
    \centering
    \begin{tabular}{|c|c|c|}
    \hline
        $\tau$ & MFPT ($\mu$s) from Eq. 13 & MFPT ($\mu$s) from eq. 12 \\ \hline
        0.02 & 1052 $\pm$ 51 & 1398 $\pm$ 11 \\ \hline
        0.05 & 4.6 $\pm$ 0.3     & 3.1 $\pm$ 0.3   \\ \hline
        0.10 & 1.1 $\pm$ 0.1     & 0.83 $\pm$ 0.02 \\ \hline
        0.50 & 0.27 $\pm$ 0.03   & 0.20 $\pm$ 0.05 \\ \hline
        1.00 & 0.10 $\pm$ 0.01   & 0.13 $\pm$ 0.01 \\
    \hline
    \end{tabular}
    \caption{Comparison between MFPTs estimated from Eq. 13 in the Main Text and from reactive flux for the double-well potential in Fig. 2. The brute force estimate is $0.14 \pm 0.08$~$\mu$s. The reactive flux estimate is obtained with Eq. 12 in the Main Text and 10000 trajectories, each one 20 ps-long . 
    }
    \label{tab:rf-vs-ole}
\end{table}

\end{document}